\documentclass[twocolumn]{jpsj2} 
\usepackage{times}

\def\runauthor{H. \textsc{Onishi} and T. \textsc{Hotta}}

\title{%
Antiferro-quadrupole state of orbital-degenerate Kondo lattice model
with \mbox{\boldmath $f^{2}$} configuration}

\author{%
Hiroaki \textsc{Onishi}$^{1,2}$
and
Takashi \textsc{Hotta}$^{1}$
}

\inst{%
$^{1}$Advanced Science Research Center, Japan Atomic Energy Agency,
Tokai, Ibaraki 319-1195, Japan \\
$^{2}$Materials Science and Technology Division, Oak Ridge National Laboratory,
Oak Ridge, TN 37831, U.S.A.
}

\recdate{October 8, 2007}

\abst{%
To clarify a key role of $f$ orbitals in the emergence of
antiferro-quadrupole structure in PrPb$_{3}$,
we investigate the ground-state property of
an orbital-degenerate Kondo lattice model
by numerical diagonalization techniques.
In PrPb$_{3}$, Pr$^{3+}$ has a $4f^{2}$ configuration and
the crystalline-electric-field ground state is
a non-Kramers doublet $\Gamma_{3}$.
In a $j$-$j$ coupling scheme,
the $\Gamma_{3}$ state is described by two local singlets,
each of which consists of two $f$ electrons
with one in $\Gamma_{7}$ and another in $\Gamma_{8}$ orbitals.
Since in a cubic structure, $\Gamma_7$ has localized nature,
while $\Gamma_8$ orbitals are rather itinerant,
we propose the orbital-degenerate Kondo lattice model
for an effective Hamiltonian of PrPb$_3$.
We show that an antiferro-orbital state is favored by
the so-called double-exchange mechanism
which is characteristic of multi-orbital systems.
}

\kword{%
PrPb$_{3}$, antiferro-quadrupole state, $j$-$j$ coupling scheme
}

\begin{document}
\maketitle


It is currently one of the central issues
in the research field of condensed-matter physics
to unveil novel magnetic phases of strongly correlated electron systems
with active orbital degrees of freedom.
It has been a common understanding that
competition and interplay among spin, charge,
and orbital degrees of freedom cause diverse ordering phenomena
involving multiple degrees of freedom,
as frequently observed in $d$- and $f$-electron systems.
\cite{Imada1998,Hotta2006}
In the case of $f$-electron systems,
spin and orbital are tightly coupled with each other
due to the strong intra-atomic spin-orbit interaction.
To describe such a complex spin-orbital state,
the $f$-electron state is usually classified
in terms of multipole degrees of freedom.

A rare-earth compound PrPb$_{3}$ with a simple AuCu$_{3}$-type cubic structure
has attracted great interest
as a typical material that exhibits antiferro-quadrupolar (AFQ) ordering.
In fact, this compound undergoes a second-order transition at 0.4~K,
\cite{Bucher1972}
which has been confirmed to be a non-magnetic but an AFQ transition.
\cite{Morin1982,Niksch1982}
In PrPb$_{3}$, Pr$^{3+}$ has a $4f^{2}$ configuration,
and the crystalline-electric-field (CEF) ground state is
a non-magnetic non-Kramers doublet $\Gamma_{3}$
with a magnetic triplet $\Gamma_{4}$ lying 19~K above the ground state.
\cite{Gross1980,Tayama2001}
The $\Gamma_{3}$ state carries $O_2^0$ and $O_2^2$ quadrupole moments.
Thus, the low-temperature property is governed by
quadrupole degrees of freedom.

Regarding the $H$-$T$ phase diagram of PrPb$_{3}$,
the so-called reentrant phase diagram has been obtained,
in which the transition temperature goes up with increasing the field
but turns to decrease and the ordered phase closes at a low field.
\cite{Tayama2001}
This reentrant behavior has been well reproduced phenomenologically
based on a mean-field theory
assuming a simple two-sublattice ordered structure.
\cite{Tayama2001}
However, recent neutron diffraction measurements have revealed that
the quadrupole ordered structure is modulated in space,
instead of a simple two-sublattice structure.
\cite{Onimaru2005}
In principle, such a long-period ordered structure could emerge
because of significant long-range quadrupole interactions,
although the origin of the long-range interactions is not clear.

So far, there have been no theoretical efforts to
understand AFQ structure of PrPb$_{3}$
from a microscopic viewpoint.
In this paper, we propose an orbital-degenerate Kondo lattice model,
which is obtained on the basis of a $j$-$j$ coupling scheme,
as an effective model for PrPb$_{3}$.
We investigate the ground-state property of the model
by using exact-diagonalization techniques.
It is found that an antiferro-orbital state emerges
due to the so-called double-exchange mechanism
which is in general relevant to multi-orbital systems.


\begin{figure}[t]
\begin{center}
\includegraphics[width=0.45\textwidth]{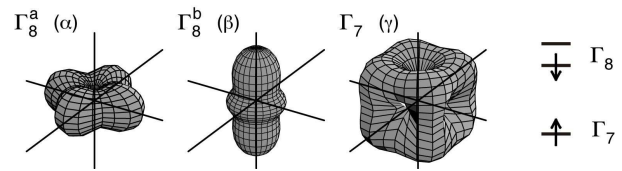}
\end{center}
\caption{%
Schematic view of $\Gamma_{8}$ and $\Gamma_{7}$ orbitals.
}
\end{figure}

First we explain the construction of an effective model for PrPb$_{3}$.
In the $j$-$j$ coupling scheme,
we first include the strong spin-orbit interaction,
and we accommodate $f$ electrons in the lower sextet
with the total angular momentum $j$=$5/2$.
Under the cubic CEF effect,
the sextet is split into a $\Gamma_{7}$ doublet and a $\Gamma_{8}$ quartet.
To distinguish two Kramers doublets in the $\Gamma_{8}$ quartet,
it is useful to introduce two orbitals,
while spin is also introduced to represent two states in each Kramers doublet.
Note that the $\Gamma_{7}$ doublet gives another orbital.
The schematic views of $\Gamma_{8}$ and $\Gamma_{7}$ orbitals are
depicted in Fig.~1.
Since we accommodate $f$ electrons in the level scheme
of the one $f$-electron state,
we refer to the level scheme of CePb$_{3}$,
which is a $4f^{1}$ compound with the same lattice structure
with that of PrPb$_{3}$.
In CePb$_{3}$, it has been found that
$\Gamma_{7}$ is the ground state and $\Gamma_{8}$ is the excited state.
\cite{Nikl1987}
Thus, for PrPb$_{3}$, we accommodate two $f$ electrons in this level scheme.

In PrPb$_{3}$, the CEF ground state is the non-Kramers doublet $\Gamma_{3}$.
In the $j$-$j$ coupling scheme, the $\Gamma_{3}$ state is described by
two local singlets, each of which is composed of two electrons
with one in $\Gamma_{7}$ and another in $\Gamma_{8}$ orbitals.
Here we note that $\Gamma_{8}$ orbitals carry
quadrupole degrees of freedom.
Taking account of the formation of local singlets,
an antiferromagnetic (AFM) coupling should be effective
between electrons in $\Gamma_{7}$ and $\Gamma_{8}$ orbitals,
although the Hund's rule coupling causes a ferromagnetic (FM) coupling.
Thus, here we introduce the AFM coupling as an effective interaction
to involve a high-order CEF effect $B_{6}^{0}$
which can not be included in the $j$=$5/2$ Hilbert space
in the $j$-$j$ coupling scheme.

Concerning the itinerancy and localized nature of orbitals,
we consider an $f$-electron hopping through the sigma bond.
In a cubic structure, due to the spatially anisotropic shape of orbital,
$\Gamma_{7}$ orbital is localized,
while $\Gamma_{8}^{a}$ orbital is itinerant in the $xy$ plane
and $\Gamma_{8}^{b}$ orbital is itinerant in all three directions.
Thus, to consider an effective model,
we assume that electron in $\Gamma_{7}$ orbital is localized,
leading to a localized spin.
Note that $\Gamma_{8}$ electron is itinerant and couples with
localized $\Gamma_{7}$ spin due to the effective AFM interaction,
which can be regarded as an analog of the Kondo coupling.

Taking into account these situations,
we obtain an orbital-degenerate Kondo lattice model
as an effective Hamiltonian for PrPb$_{3}$, given by
\begin{align}
 H=
 &
 \sum_{\langle {\bf i},{\bf j} \rangle,\tau,\tau'\sigma}
 t_{\tau\tau'}^{{\bf i}-{\bf j}}
 f_{{\bf i}\tau\sigma}^{\dag} f_{{\bf j}\tau'\sigma}
 +J_{\rm K}
 \sum_{{\bf i}}{\bf S}_{{\bf i}\Gamma_7}\cdot{\bf S}_{{\bf i}\Gamma_8}
\nonumber\\
 &
 +U\sum_{{\bf i},\tau}\rho_{{\bf i}\tau\uparrow}\rho_{{\bf i}\tau\downarrow}
 +U'\sum_{{\bf i}}\rho_{{\bf i}\alpha}\rho_{{\bf i}\beta}
\nonumber\\
 &
 +J\sum_{{\bf i},\sigma,\sigma',\tau\neq\tau'}
  f_{{\bf i}\tau\sigma}^{\dag}f_{{\bf i}\tau'\sigma'}^{\dag}
  f_{{\bf i}\tau\sigma'}f_{{\bf i}\tau'\sigma}
\nonumber\\
 &
 +J'\sum_{{\bf i},\sigma\neq\sigma',\tau\neq\tau'}
  f_{{\bf i}\tau\sigma}^{\dag}f_{{\bf i}\tau\sigma'}^{\dag}
  f_{{\bf i}\tau'\sigma'}f_{{\bf i}\tau'\sigma}
\end{align}
where $f_{{\bf i}\tau\sigma}$ is the annihilation operator
for $\Gamma_{8}$ electron
with spin $\sigma$(=$\uparrow,\downarrow$)
in orbital $\tau$(=$\alpha,\beta$) at site ${\bf i}$,
$\rho_{{\bf i}\tau\sigma}$=$f_{{\bf i}\tau\sigma}^{\dag}f_{{\bf i}\tau\sigma}$,
$\rho_{{\bf i}\tau}$=$\sum_{\sigma}\rho_{{\bf i}\tau\sigma}$,
${\bf S}_{{\bf i}\Gamma_8}$=%
$(1/2)\sum_{\sigma\sigma'\tau}
 f_{{\bf i}\tau\sigma}^{\dag}
 \mbox{\boldmath $\sigma$}_{\sigma\sigma'}
 f_{{\bf i}\tau\sigma'}$,
where $\mbox{\boldmath $\sigma$}_{\sigma\sigma'}$ are Pauli matrices,
and ${\bf S}_{{\bf i}\Gamma_7}$ is the spin-1/2 operator for $\Gamma_{7}$ spin.
The summation of $\langle {\bf i},{\bf j} \rangle$ is taken for
nearest neighbor sites in the cubic lattice.
The hopping amplitudes are evaluated from the overlap integral
between $f$-orbital wavefunctions in adjacent sites,
which are given by
$t_{\alpha\alpha}^{{\bf x}}$=$3t/4$,
$t_{\alpha\beta}^{{\bf x}}$=%
$t_{\beta\alpha}^{{\bf x}}$=$-\sqrt{3}t/4$,
$t_{\beta\beta}^{{\bf x}}$=$t/4$
for the $x$ direction,
$t_{\alpha\alpha}^{{\bf y}}$=$3t/4$,
$t_{\alpha\beta}^{{\bf y}}$=%
$t_{\beta\alpha}^{{\bf y}}$=$\sqrt{3}t/4$,
$t_{\beta\beta}^{{\bf y}}$=$t/4$
for the $y$ direction, and
$t_{\beta\beta}^{{\bf z}}$=$t$,
$t_{\alpha\alpha}^{{\bf z}}$=%
$t_{\alpha\beta}^{{\bf z}}$=%
$t_{\beta\alpha}^{{\bf z}}$=$0$
for the $z$ direction,
where $t$=$(3/7)(ff\sigma)$.
Hereafter, $t$ is taken as the energy unit.
In the second term,
$J_{\rm K}$ is the Kondo coupling
between $\Gamma_{7}$ spin and $\Gamma_{8}$ electron.
The rest terms are interactions among $\Gamma_{8}$ electrons:
$U$, $U'$, $J$, and $J'$ denote
intra-orbital,
inter-orbital,
exchange,
and pair-hopping interactions, respectively.
Note that the relation $U$=$U'$+$J$+$J'$ holds,
which originates from the rotational invariance in the orbital space,
and $J$=$J'$ is assumed.
\cite{Hotta2006}

We analyze the model (1) by numerical diagonalization.
Since the size of the Hilbert space becomes so large as $32^{N}$
due to orbital degree of freedom, where $N$ is the number of sites,
it is rather difficult to enlarge the system size.
However, the method is advantageous to grasp the ground-state property,
such as orbital structure, in an unbiased manner.
In the present work,
first we study a $2$$\times$$2$ square four-site system in the $xy$ plane.
Then, taking account of the characteristics
grasped within the four-site system,
we proceed to a $2$$\times$$2$$\times$$2$ cubic eight-site system.
In this paper, we set $U'/W$=$2$, where $W$ is the band width,
and investigate the dependence on $J_{\rm K}$ and $J$.
Note that the band width is $W$=$4$ for the square system
and $W$=$6$ for the cubic system.


\begin{figure}[t]
\begin{center}
\includegraphics[width=0.48\textwidth]{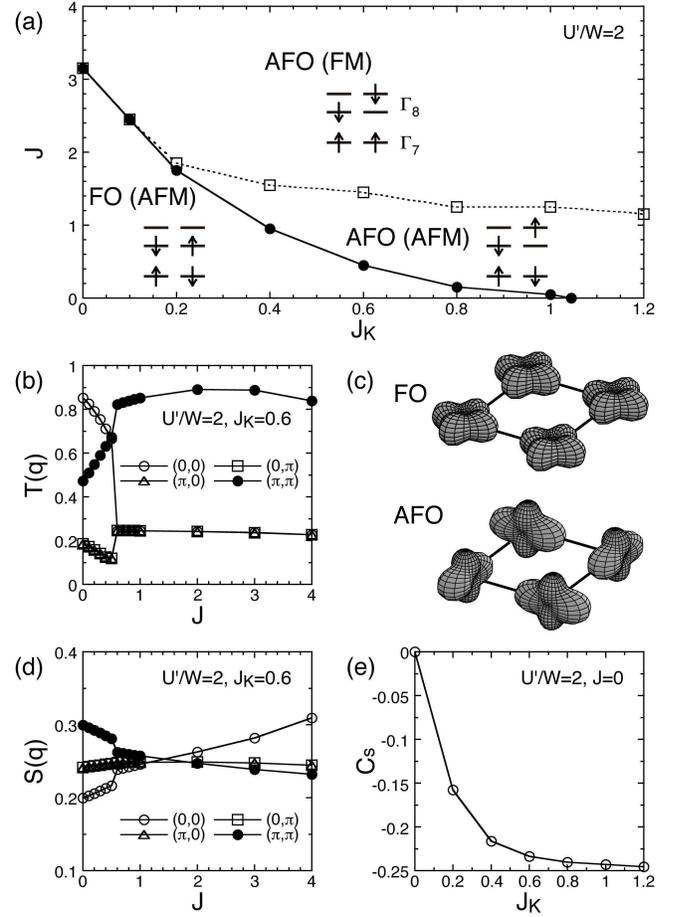}
\end{center}
\caption{%
Four-site results.
(a) Ground-state phase diagram in the $(J_{\rm K},J)$ plane.
Inset denotes schematic view of electron configuration in each phase.
(b) $T({\bf q})$ as a function of $J$ at $J_{\rm K}$=$0.6$.
(c) FO and AFO structures.
(d) $S({\bf q})$ as a function of $J$ at $J_{\rm K}$=$0.6$.
(e) $C_{s}$ as a function of $J_{\rm K}$ at $J$=$0$.
Note that $C_{\rm s}({\rm i})$ takes the equivalent value at every site
due to the translational symmetry.
}
\end{figure}

First, we show the results for the four-site system.
The main result is summarized in Fig.~2(a),
which is the ground-state phase diagram in the $(J_{\rm K},J)$ plane.
There three types of competing spin-orbital configurations are observed.
When $J_{\rm K}$ and $J$ are small,
we find a ferro-orbital (FO) state
with an AFM configuration
in each of $\Gamma_{7}$ and $\Gamma_{8}$ orbitals,
while spins in $\Gamma_{7}$ and $\Gamma_{8}$ orbitals are
antiparallel at each site due to $J_{\rm K}$.
With increasing $J$,
an antiferro-orbital (AFO) state occurs,
and the spin state turns to be FM at a larger $J$
than the transition point from FO to AFO.
It should be noted that even though the spin state is characterized by
AFM or FM in each of $\Gamma_{7}$ and $\Gamma_{8}$ orbitals,
a local singlet is formed due to $J_{\rm K}$
and the ground state is totally non-magnetic.

Let us here discuss the orbital state.
In order to determine the orbital structure,
it is useful to introduce new operators for $\xi$ and $\eta$ orbitals,
which are given by linear combinations of
the original operators, such as
\begin{align}
 \tilde{f}_{{\bf i}\xi\sigma}=
 &
 \cos(\theta_{\bf i}/2)f_{{\bf i}\beta\sigma}+
 \sin(\theta_{\bf i}/2)f_{{\bf i}\alpha\sigma},
\nonumber\\
 \tilde{f}_{{\bf i}\eta\sigma}=
 &
 -\sin(\theta_{\bf i}/2)f_{{\bf i}\beta\sigma}+
  \cos(\theta_{\bf i}/2)f_{{\bf i}\alpha\sigma},
\end{align}
where $\theta_{\bf i}$ characterizes the orbital shape at each site.
The optimal $\{\theta_{\bf i}\}$ is determined so as to
maximize the orbital structure factor, defined by
\begin{equation}
 T({\bf q})=
 \sum_{{\bf j},{\bf k}}
 \langle \tilde{T}_{\bf j}^{z}\tilde{T}_{\bf k}^{z} \rangle
 e^{{\rm i}{\bf q}\cdot({\bf j}-{\bf k})}/N,
\end{equation}
where
$\tilde{T}_{\bf i}^z$=$\sum_{\sigma}
(\tilde{f}_{{\bf i}\xi\sigma}^{\dag}
 \tilde{f}_{{\bf i}\xi\sigma}
 $$-$$
 \tilde{f}_{{\bf i}\eta\sigma}^{\dag}
 \tilde{f}_{{\bf i}\eta\sigma})/2$
and $\langle \cdots \rangle$ denotes the expectation value.
In Fig.~2(b),
we show $T({\bf q})$ as a function of $J$ at $J_{\rm K}$=$0.6$
with $\theta_{\bf i}$=$\theta$.
For small $J$, the dominant component is $T(0,0)$ with $\theta$=$\pi$,
indicating FO state.
In the FO state of the present square system in the $xy$ plane,
$\Gamma_{8}^{a}$ orbitals are favorably occupied to gain kinetic energy,
since $\Gamma_{8}^{a}$ orbital extends to adjacent sites,
as shown in Fig.~2(c).
With increasing $J$,
the dominant component changes to $T(\pi,\pi)$ with $\theta$=$\pi/2$,
indicating AFO state in Fig.~2(c).
We note that the present square lattice structure is reflected
in the orbital shape.
In this context, it is crucial to proceed to the eight-site system
to discuss the orbital structure in the cubic lattice.

Concerning the spin state,
we measure the spin structure factor of $\Gamma_{8}$ electrons,
defined by
\begin{equation}
 S({\bf q})=
 \sum_{{\bf j},{\bf k}}
 \langle S_{{\bf j}\Gamma_8}^z S_{{\bf k}\Gamma_8}^z \rangle
 e^{{\rm i}{\bf q}\cdot({\bf j}-{\bf k})}/N.
\end{equation}
As shown in Fig.~2(d),
with increasing $J$, $S(0,0)$ is increased, while $S(\pi,\pi)$ is reduced,
so that the dominant spin correlation changes from AFM to FM.
Here we note that the motion of $\Gamma_{8}$ electrons leads to
a FM spin arrangement.
Namely, the double-exchange mechanism is effective,
which is characteristics of multi-orbital systems.

We also measure the on-site spin correlation,
defined by
\begin{equation}
 C_{\rm s}({\bf i})=
 \langle S_{{\bf i}\Gamma_7}^z S_{{\bf i}\Gamma_8}^z \rangle.
\end{equation}
In Fig.~2(e),
the $J_{\rm K}$ dependence of $C_{\rm s}$ at $J$=$0$ is shown.
It is obvious that $C_{\rm s}$=$0$ at $J_{\rm K}$=$0$,
since there is no correlation
between $\Gamma_{7}$ spin and $\Gamma_{8}$ electron.
With increasing $J_{\rm K}$,
$C_{\rm s}$ decreases and gradually approaches $-$$1/4$,
indicating the stabilization of the local singlet.
It is found that even when $J$ is increased,
$C_{\rm s}$ keeps a value near $-$$1/4$ (not shown),
indicating the robust formation of the local singlet.
Here it is worth noting that
as $J_{\rm K}$ increases and the local singlet is stabilized,
the AFO phase tends to extend to the region of small $J$,
as shown in Fig.~2(a).
Namely, the double-exchange mechanism becomes significant due to $J_{\rm K}$.


\begin{figure}[t]
\begin{center}
\includegraphics[width=0.48\textwidth]{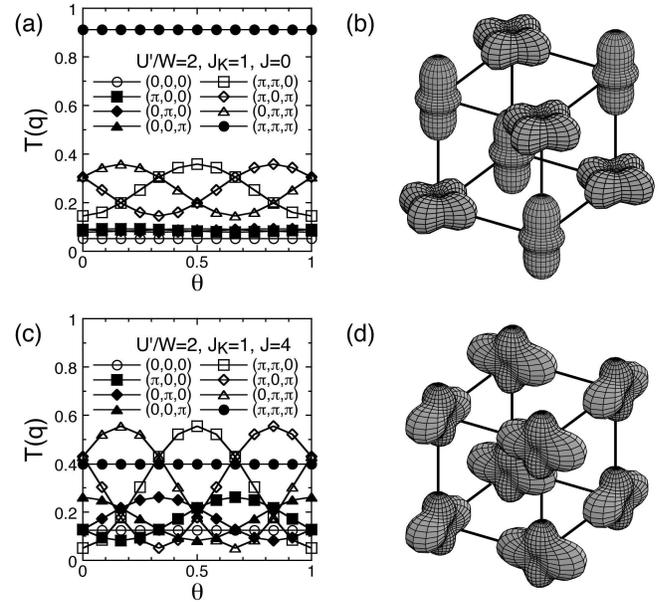}
\end{center}
\caption{%
Eight-site results.
(a) $T({\bf q})$ as a function of $\theta$ and
(b) orbital structure at $J_{\rm K}$=$1$ and $J$=$0$.
(c) $T({\bf q})$ as a function of $\theta$ and
(d) orbital structure at $J_{\rm K}$=$1$ and $J$=$4$.
}
\end{figure}

Now we move on to the results for the eight-site system.
In Fig.~3(a), we show the $\theta$ dependence of $T({\bf q})$
at $J_{\rm K}$=$1$ and $J$=$0$.
We find that $T(\pi,\pi,\pi)$ is dominant, indicating AFO state.
Concerning the orbital shape,
it is observed that the magnitude of $T(\pi,\pi,\pi)$
does not depend on $\theta$.
We can not determine the actual orbital shape,
but the AFO structure with any $\theta$ is possible to realize.
The AFO structure with $\theta$=$0$ is shown in Fig.~3(b).
As for the spin state, $S(\pi,\pi,\pi)$ is found to be dominant.
We note that the present parameter set ($J_{\rm K}$=$1$ and $J$=$0$)
is corresponding to the FO phase
in the four-site system, as shown in Fig.~2(a).
It is naively expected that the AFO phase extends to a broad area
in the phase diagram even when we consider the cubic lattice.
However, at $J_{\rm K}$=$1$ and $J$=$4$,
$T({\bf q})$ has dominant components
$T(\pi,\pi,0)$ with $\theta$=$\pi/2$,
$T(\pi,0,\pi)$ with $\theta$=$5\pi/6$,
and $T(0,\pi,\pi)$ with $\theta$=$\pi/6$,
as shown in Fig.~3(c),
while $S(0,0,0)$ is dominant.
The $(\pi,\pi,0)$ orbital structure is depicted in Fig.~3(d).
When $J$ is further increased,
the orbital structure is considered to turn to be AFO
due to the double-exchange mechanism,
suggesting a rich phase diagram including competing orbital states.

Finally, we briefly discuss possible relevance of the present results
to the AFQ structure of PrPb$_{3}$.
We have shown that for a one-dimensional $j$-$j$ coupling model
with an $f^{1}$ configuration,
\cite{Onishi2006}
an incommensurate orbital state appears
due to the competition between itinerant and localized orbitals.
By analogy, we expect that the competition among plural orbital states
with different nature could cause a modulated orbital structure
as observed in PrPb$_{3}$.


In summary,
we have investigated the ground-state property of
the orbital-degenerate Kondo lattice model to understand
the quadrupole structure in PrPb$_{3}$ from a microscopic viewpoint.
We have observed several types of competing spin-orbital states.
In particular,
it has been emphasized that
the AFO state emerges due to the double-exchange mechanism.


The authors have been supported by a Grant-in-Aid for
Scientific Research in Priority Area ``Skutterudites''
under the contract No.~18027016
from the Ministry of Education, Culture, Sports, Science, and
Technology of Japan.
T.H. has been also supported by a Grant-in-Aid for
Scientific Research (C)
under the contract No.~18540361
from Japan Society for the Promotion of Science.


\end{document}